\documentclass[11pt]{article}

\usepackage[]{graphicx,float,latexsym,times}
\usepackage{amsfonts,amstext,amsmath,amssymb,amsthm}
\usepackage{mathrsfs}

\usepackage{caption2}

\long\def\symbolfootnote[#1]#2{\begingroup%
\def\thefootnote{\fnsymbol{footnote}}\footnote[#1]{#2}\endgroup}
\usepackage{titlesec} % Very fancy section title control: Sets sections as SQA requires
\titleformat{\section}{\large\bfseries}{\thesection.}{.5em}{}
\titlespacing*{\section}{0pt}{*3}{*2}
\titleformat{\subsection}{\normalfont\bfseries}{\thesubsection.}{.5em}{}
\titlespacing*{\subsection} {0pt}{*3}{*2}
\titleformat{\subsubsection}{\normalfont\bfseries}{\thesubsubsection.}{.5em}{}
\titlespacing*{\subsubsection} {0pt}{*3}{*2}

\theoremstyle{plain} %% italic text
\newtheorem{theorem}{Theorem}[section]
\newtheorem{lemma}{Lemma}[section]
\newtheorem{corollary}{Corollary}[section]

\theoremstyle{definition} %% or \theoremstyle{remark} will produce roman text

\newtheorem{remark}{Remark}[section]

\numberwithin{equation}{section} %% double numbering within sections

\newcommand{\bs}{\boldsymbol}
\newcommand{\D}{\displaystyle}
\newcommand{\mf}{\mathfrak}
\newcommand{\mscr}{\mathscr}
\newcommand{\mc}{\mathcal}

\usepackage[authoryear]{natbib}
\begin{document}

\title{\textbf{\Large Optimal Group-Sequential Tests with Groups 
of Random Size}}

\date{}

\maketitle

%%%%%%%%% Authors, affiliations %%%%%%%%%%%%%%%%%%%%%%%%%%

\author{
\begin{center}
\vskip -1cm

\textbf{\large A. Novikov, X.I. Popoca Jim\'enez  }

Department of Mathematics, Metropolitan Autonomous University - Iztapalapa,\\
 Mexico City, Mexico

\end{center}
}

\symbolfootnote[0]{\normalsize Address correspondence to X.I. Popoca-Jiménez, Departamento de Matem\'aticas, UAM-Iztapalapa, San Rafael Atlixco 186, col. 
Vicentina, C.P. 09340, Mexico City, Mexico; E-mail: dextera80@yahoo.es}

{\small \noindent\textbf{Abstract:} We consider sequential hypothesis testing based on observations which are
received in groups of random size. The observations are
assumed to be independent both within and between the groups. We assume 
that the group sizes are independent and  their distributions are known,  
and that
the groups are formed independently of the observations.   

We are concerned with a problem of  testing a simple hypothesis  against a
simple alternative.  For any (group-) sequential test, 
we take into account the following three characteristics: its type I and 
type II error probabilities and the average cost
of observations.
 Under  mild conditions,
we characterize the structure of  sequential tests 
minimizing the average cost of observations among  all sequential
tests whose type I and type II error probabilities
do not exceed some prescribed levels.}
\\ \\
%%%%%%%%% Key words %%%%%%%%%%%%%%%%%%%%%%%%%%
{\small \noindent\textbf{Keywords:} Sequential analysis;
Hypothesis testing;
Optimal stopping;
Optimal sequential tests}
\\ \\
%%%%%%%%% Subject Classifications %%%%%%%%%
{\small \noindent\textbf{Subject Classifications:} 62L10, 62L15, 62F03, 60G40.}

\section{Introduction}

In this article, we consider sequential hypothesis testing when the
observations are received in groups of a random size, rather than
on a one-at-a-time basis (we adhere to the statistical model proposed by
%N.~Mukhopadhyay and B.~M.~de~Silva
\cite{Mukhopadhyay} for this context).  There are many practical 
situations where the
random group size model comes
into question and a lot of theoretical problems that arise
(see \cite{Mukhopadhyay}). In this article we address the problems of 
optimality 
of the random group-sequential tests for the case of two simple hypotheses, 
covering theoretical 
aspects of their optimality.

In the case of random groups of observations, there are different ways to
quantify the volume of observations taken for the analysis; e.~g., a
special interest can be put on the total number of observations, or on the
number of groups taken (see \cite{Mukhopadhyay}). To tackle the
possible differences, in this article we  introduce a natural concept of 
cost of observations
accounting  for the  number of groups and/or for the number of
observations within the groups, and use the average cost as
one of the  characteristics to be taken into account. 

Our main objective is to characterize all the tests  minimizing the average 
cost of the experiment, among all
group-sequential tests whose type I and type II error probabilities do not
exceed some given levels.

With respect to these, we should start with the classical framework  
of one-per-group observations where the average sample number is minimized
under restrictions on the probabilities of the first and the second kind. 
 \cite{waldwolfowitz}  show that Wald's Sequential Probability Ratio 
Test (SPRT) has a minimum average sample number, among all the tests 
whose error probabilities do not exceed those of the SPRT. The minimum is 
reached both under the null-hypothesis and under the alternative - this 
strong optimality property is known as the Wald-Wolfowitz optimality. 

For the group-sequential model we 
adhere 
to in this article,  \cite{Mukhopadhyay} proposed 
an extension of the classical SPRT, called RSPRT. In this article, we want 
to characterize the structure of optimal sequential  tests and, in 
particular, show the optimality of the RSPRT, in the Wald-Wolfowitz sense,  when the group sizes are identically distributed.

For a more general case, when the group sizes not necessarily have the same distribution, we use a weaker approach related to the 
minimization of the average cost
under {\em one } hypothesis (see \cite{Lorden}). If an optimal, in the Wald-Wolfowitz sense, test will 
ever 
be found, for any specific group size distribution model, it  should minimize the average cost 
{\em 
each one} of the hypothesis, thus it should be of the particular form we 
 find here.

In this way, our main concern in this article is the characterization of optimal group-sequential tests, which minimize the average cost  of the observations given restrictions on the error probabilities.

In Section \ref{sec2}, the main definitions and assumptions are presented.

In Section \ref{sec3}, the  problem of findinig optimal test  is reduced to an optimal stopping problem.

In Section \ref{a20}, characterizations of optimal sequential rules are given.

In Section \ref{sec5}, the optimality  of the random sequential probability ratio tests (RSPRT)  is demonstrated.

The proofs of the main results are placed in the Appendix A.  

\section{Notation and Assumptions}\label{sec2}

We assume that i.i.d.  observations  $X_{kj}$,  $j = 1, 2,
\ldots, n_k$, are available to the statistician sequentially, in groups
numbered by $k = 1, 2, \ldots$. The group sizes $n_k$ are assumed to
be values of some independent integer-valued random variables $\nu_k$.
%in such a way that  $X_{kj}$,  $j=1,2,\dots,n_k$ are independent and identically distributed under  condition that
%$\nu_1=n_1, \nu_2=n_2,\dots, \nu_k=n_k$, $(n_1,n_2,\dots,n_k)\in \mathscr G^k$, $k=1,2,\dots$, where $\mathscr G$ is the set of possible group sizes.
The distributions of $\nu_k$ are assumed to be fixed and known to the
statistician. But the distribution of $X_{kj}$ (we denote it $P_\theta$)
depends on a parameter 
$\theta$, and the goal of the statistician is to test two simple 
hypothesis, 
$H_{0}:\theta=\theta_0$ against  $H_1:
\theta=\theta_1$. The number of groups to be taken for the analysis is up 
to
the statistician,
and is to be determined on the basis of observations he
 has up to the moment of stopping.

Below we formalize  this procedure in detail.

For any natural $k$, we denote by $X_k^{(n_k)}$ the vector of
observations in the $k$-th group, $X_k^{(n_k)} =(X_{k1}, \ldots, 
X_{kn_k})$. If
$n_k=0$ (no observations in the group), we will formally write it as $()$.
Group sizes $n_1,\dots,n_k$ are assumed to be values of
independent variables $\nu_1,\dots,\nu_k$, with 
respective probability mass functions (p.m.f.) 
\begin{equation}
 P(\nu_i=n)=p_i(n),\quad n\in \mathscr G \subset \{0,1,2,\dots\},
\end{equation}
$i=1,2,\dots,k$; $k=1,2,\dots$
Then the joint distribution of the first $k$ consecutive group sizes (let 
us denote $ \nu^{(k)}=(\nu_1,\dots,\nu_k)$ and $n=(n_1, \dots, n_k) \in \mathscr G^k$ ) is given by
$$P(\nu^{(k)}= n)=p( n) = \prod_{i=1}^k p_i(n_i),\,{ n}\in\mathscr 
G^k, $$
 for all $k=1, 2, \dots$

It is assumed that the observations $X_{ni}$,  $i=1,2,\dots,n_k$, $n=1,2,\dots, k$  
are conditionally independent and identically distributed, given
group sizes  $\nu_1=n_1,\nu_2=n_2,\dots, \nu_k=n_k$, for all $n=(n_1,\dots,n_k)\in\mathscr G^ k$, $k=1,2,\dots$

Let  $X_{ni}$ take its  ``values'' in a measurable space  
$(\mathfrak
X,\mathscr X)$, and assume  that its distribution $P_\theta$ has  a ``density''
function
$f_\theta$ on $\mathfrak X$ (the Radon--Nikodym derivative of $P_\theta$ 
with
respect to a $\sigma$-finite measure $\mu$ on $\mathscr X$).

In this way, for any number of groups $k=1,2,\ldots$, given any 
consecutive group
sizes  ${n}=(n_1,\ldots,n_k)\in\mathscr G^k$ the random
vector of observations
$ X^{ (n)}=\left(X_1^{(n_1)}, \ldots, X_k^{(n_k)}\right)$
has a joint density
$$ f_\theta^{(n)}(x^{( n)})=f_\theta^{( n)}\left(x_1^{(n_1)}, 
\ldots, 
x_k^{(n_k)}\right)
= \prod_{i=1}^k \prod_{j=1}^{n_k} f_\theta(x_{ij}) $$
with respect to the product measure
$ \mu^{ n} = \mu^{n_1} \otimes \cdots \otimes \mu^{n_k}, $
on $\mathscr X^{ n}=\mathscr X^{n_1}\otimes\cdots\otimes \mathscr 
X^{n_k}$,
where
$ \mu^{n_k} = \mu\otimes\cdots\otimes\mu$
($n_k$ times).
By definition, we assume here that $\prod_{j=1}^0(\cdot)\equiv 1$
and that $\mu^0$ is a probability  measure on the (trivial) 
$\sigma$-algebra
$\mathscr X^0$ on $\mathfrak X^0=\{()\}$.

Throughout the paper, it will be assumed that the distributions $P_{\theta_0}$ 
and 
$P_{\theta_1}$ are distinct:
\begin{equation}\label{09-1115}
\mu\{x:f_{\theta_0}(x)\not=f_{\theta_1}(x)\}>0. 
\end{equation}

We define a (randomized) {\em stopping rule} $\psi$ as a family of
measurable functions
$
\psi_{ n}:\mathfrak X^{ n}\mapsto [0,1]
$, ${ n}\in {\mathscr G}^k$, $k\geq 1$,
where $\psi_{ n}(x^{ (n)})$
represents the conditional probability to stop,  given number of groups 
$k$, group
sizes $ n=(n_1,\ldots,n_k)$, and the data $x^{(n)}$ observed till the time 
of stopping. 

In a similar manner, a (randomized) {\em  decision rule} $\phi$ is a 
family 
of
measurable functions
$
\phi_{ n}:\mathfrak X^{ n}\mapsto [0,1]
$, ${ n}\in {\mathscr G}^k$,  $k\geq 1$,
where $\phi_{ n}(x^{( n)})$
represents the conditional probability to {\em 
reject } $H_0$, given a number $k$ of the groups observed, group
sizes $n=(n_1, \dots, n_k)$ and  data $x^{ (n)}$ observed till the time 
of final decision (to accept or reject $H_0$).

A group-sequential test is a 
pair
$(\psi,\phi)$ of a stopping rule $\psi$ and a decision rule $\phi$.

For ${ n}=(n_1,\dots,n_k)\in\mathscr G^k$, with any $k\geq 1$, let
\begin{equation}\label{31-1252}
t_{ n}^\psi  \big(x^{ (n)}\big) =
\left(1 - \psi_{(n_1)}(x_1^{(n_1)})\right)
\cdots
\left(1 - \psi_{(n_1,
\ldots,n_{k-1})}(x_1^{(n_1)},\ldots,x_{k-1}^{(n_{k-1})})\right),
\end{equation}
\begin{equation}\label{31-1253}
s_{ n}^\psi(x^{( n)}) =
t_{ n}^\psi(x^{( n)})
  \psi_{ n}(x^{ (n)})
\end{equation}
(by definition, $t_{ n}^\psi(x^{ (n)}) \equiv 1$ for 
${ n}\in \mathscr
G$).

Let us also denote for $n\in \mathscr G^k$, $k\geq 1$
$$S_{n}^\psi=\left\{x^{(n)}\in \mathfrak X^{n}: \,s_{n}^\psi(x^{(n)})>0\right\}\; \text{ and } \;
T_{n}^\psi=\left\{x^{(n)} \in \mathfrak X^{n}:t_{n}^\psi(x^{ 
	(n)})>0\right\}.$$

In the latter expression, we assume
$ s_{ n}^\psi = s_{ n}^\psi\left(X^{(n)}\right) 
$, in spite of its initial definition as
$ s_{ n}^\psi(x^{(n)}) $.
Throughout the paper, we will use this kind of double interpretation for 
any function of 
observations, according to  the following rule. 
If $F_{n}$, ${ n}\in\mathscr G^k$,
is some function of observations
%($F_{ n} = F_{ n}(x_1^{(n_1)}, x_2^{(n_2)}, \ldots, x_k^{(n_k)})$
%or $F_{ n} = F_{ n}(X_1^{(n_1)}, X_2^{(n_2)}, \ldots, X_k^{(n_k)})$)
and its arguments are omitted, then $ F_{ n}$ is interpreted as $ 
F_{ 
	n}(X^{(n)}) $
when it is under the probability or expectation sign, and
as
$F_{ n}(x^{(n)}) $, otherwise.

{Any stopping rule $\psi $ generates a random 
variable $\tau_{\psi}$ (stopping time), with a p.m.f.
$$P_\theta(\tau_{\psi} = k) = \sum_{{ n}\in\mathscr G^k} p({ n}) E_{\theta}
s_{ n}^\psi, \quad k=1,2,\ldots,$$
where
$E_\theta(\cdot)$ is expectation with respect to $P_\theta$.
Under $H_i$,
a stopping rule  $\psi$  terminates 
the testing procedure with probability 1 if 
\begin{equation}\label{And7}
P_{\theta_i}(\tau_{\psi}<\infty)=
\sum_{k=1}^\infty\sum_{{ n}\in\mathscr G^k} p({ n}) E_{\theta_i}
s_{ n}^\psi =1.
\end{equation}
Let us denote $\mathscr F_i $ the set of stopping rules satisfying
\eqref{And7},  and let  $\mathscr S_i$ be the set of all 
group-sequential tests $(\psi,\phi) $ 
such that $\psi\in \mathscr F_i$, $i=0,1$.

For brevity, we will write $E_{i}$, $f_{i}$ and $P_{i}$ instead of 
$E_{\theta_{i}}$, $f_{\theta_{i}}$ and $P_{\theta_{i}}$, respectively,
$i=0,1$. \\
%Let us define 
%\begin{equation}\label{And8}
%\mathscr F=\{\psi: P_0(\tau_\psi<\infty)=1,\,P_1(\tau_\psi<\infty)=1   \},
%\mathscr F=\{\psi:\sum_{n\in \mathscr G^k}p(n) 
%E_0t_n^\psi\to 0,\;\sum_{n\in \mathscr G^k}p(n) 
%E_1t_n^\psi\to 0,\; k\to\infty\}.
%\,P_1(\tau_\psi<\infty)=1   \}
%\end{equation}
%It follows from Lemma \ref{And1-1} in the Appendix that $\mathscr F$ is 
%the class of stopping rules $\psi$ for which (\ref{And7}) is satisfied. 

For a group-sequential test $(\psi, \phi) $ , the type I and 
type II error probabilities are defined as
\begin{eqnarray*}
\alpha(\psi, \phi) &=& P_{0}\big(\mbox{reject}\,H_0\,\mbox{using}\,(\psi,\phi)\big) =
\sum_{k=1}^\infty \sum_{{ n} \in\mathscr G^k} p({ n}) E_{0} s^\psi_{ 
n}
\phi_{ n}, \quad \mbox{and}\\
\beta(\psi, \phi) &=& P_{1}\big(\mbox{accept}\,H_0 \,\mbox{using}\,(\psi,\phi)\big) =
\sum_{k=1}^\infty \sum_{{ n} \in\mathscr G^k} p({ n}) E_{1} s^\psi_{ 
n}
(1-\phi_{ n}).
\end{eqnarray*}

Let us assume that the cost of  $m$ observations  is 
$c(m)> 0$,  $m\in\mathscr G$.
The cost of the first $k$ stages of the experiment for a given $n \in 
\mathscr 
G^k$ is then
$ c( n) = \sum_{i=1}^k c(n_i).$
The average cost of obtaining the $k$-th group of observations is
$$ \bar c_k =\sum_{n\in \mathscr G} p_k(n) c(n).$$

We assume that  $0<\bar{c}_k <\infty$  for all $k\geq 1$. The average total 
cost of the sequential sampling according to a stopping  
rule 
 $\psi \in \mathscr{F}_i$ is, {under}
$ H_i,$
$$ K_i(\psi) = \sum_{k=1}^\infty\sum_{ n\in\mathscr G^k}p(n)c({ n})
E_{i} s^\psi_{ n}, \,i=0,1. $$
Here are some natural particular cases of the cost structure: if
$c(m) = 1$, for all $m\in \mathscr G$,  then $K_i(\psi)$ corresponds to the  average
number of the groups taken;
another particular case is
$ c(m)=m$ for all $m\in \mathscr G$, it accounts for the average  total number of
observations. A combination of these two seems to be quite useful  for some applications: 
$c(m)=a+bm$, with $a, b >0$ (see, for example, \cite{schmitz}).

It is seen from the above definitions that our group-sequential 
experiment always starts 
with one group of observations (i.e., stage $k=1$ is always present). 
The usual (for the sequential analysis) case when the experiment admits 
stopping without taking  observations  can be easily 
 incorporated in this scheme by supposing that the first group always has 
size $n_{1}=0$ (and stopping at this stage means no observations will be 
taken).

The usual context for sequential hypothesis testing is  to minimize average costs under restrictions on the type I and type II
error probabilities. We are concerned in this article with minimizing $K_0(\psi)$ and/or $K_1(\psi)$ under  restrictions
\begin{equation}\label{26-1833}
 \alpha(\psi,\phi)\leq \alpha\quad \mbox{and} \quad \beta(\psi,\phi)\leq \beta,
\end{equation}
over $(\psi,\phi)\in\mathscr S_0$ (and/or $(\psi,\phi)\in\mathscr S_1$), where $\alpha\in[0,1]$ and $\beta\in[0,1]$ are some given numbers.

\section{Reduction to optimal stopping problem}
\label{sec3}

The problem of minimizing average cost $K_0(\psi)$ under restrictions on the error probabilities $\alpha (\psi,\phi)\leq \alpha$ and $\beta (\psi,\phi)\leq \beta$ is routinely reduced to a non-constrained optimization problem using the Lagrange multipliers method. 

Let us define  
the Lagrangian function
\begin{equation}\label{Reduction.Eq1}
 L(\psi, \phi; \lambda_{0}, \lambda_{1}) :=  K_0(\psi) + 
\lambda_{0}\alpha(\psi, \phi) + \lambda_{1} \beta(\psi, \phi),
\end{equation}
where $\lambda_{0}>0$ and $\lambda_{1}>0$ are constant multipliers.

The following Lemma is the essence  of the reduction, and is almost trivial. 
 It is placed here  for convenience of references.
\begin{lemma}\label{a-l1} 
Let  $\lambda_{0},\lambda_{1}>0$ and a test
$(\psi, \phi)\in  \mathscr S_0$ be such that
\begin{equation}\label{a6} L(\psi, \phi;\lambda_{0},\lambda_{1}) 
\leq  
L(\psi^\prime, 
\phi^\prime;\lambda_{0},\lambda_{1})\end{equation}
for all $(\psi^\prime, \phi^\prime)\in \mathscr S_0$.
Then for every test $(\psi^\prime, \phi^\prime)\in\mathscr S_0$  such that
\begin{equation}\label{Reduction.Eq4}
\alpha(\psi^\prime, \phi^\prime) \leq \alpha(\psi, \phi)  \quad{and}\quad 
\beta(\psi^\prime, \phi^\prime) \leq \beta(\psi, \phi),
\end{equation}
it holds
\begin{equation}\label{Reduction.Eq5}
 K_0(\psi^\prime) \geq K_0(\psi).
\end{equation}
The inequality in (\ref{Reduction.Eq5}) is strict if at least one of the 
inequalities in (\ref{Reduction.Eq4}) is strict.
\end{lemma}

Let us define $g(z) = g(z;\lambda_{0},\lambda_{1})=\min\{\lambda_{0}, 
\lambda_{1}z\}$ for all  $0\leq 
z<\infty$,   and let 
$\mathcal{P}_{k} =\{{ n} \in \mathscr{G}^{k}: p({ n})>0\}$, for 
$k\geq 1$.

For all  $k\geq 1$ and  all  ${ n}\in \mathscr G^k $ let
$$ z_{ n} = z_{ n}\left(x^{(n)}\right) 
=\begin{cases}\frac{f_{1}^{{(n)}}(x^{{( n)}})}{f_{0}^{{( n)}}(x^{{( 
n)}})},& \mbox{if}\quad  f_{0}^{{(n)}}(x^{{( n)}})>0, \\
\; \;\infty,& \mbox{if}\; f_{0}^{{(n)}}(x^{{( n)}})=0,\;\mbox{ but}\; 
f_{1}^{{(n)}}(x^{{( n)}})> 0,\\
\;\; 0\; (\mbox{or whatever}),& \;\mbox{ otherwise.}
\end{cases}$$

From this time on, we will use the following notation.

Let us write $\psi\simeq 
I_{\{F_1\preccurlyeq F_2\}}$ (say)
when
$
 I_{\{F_1< F_2\}}\leq \psi\leq  I_{\{F_1\leq F_2\}}.
$
It is easy to see that this is an equivalent way to say that:
%\begin{tabular}{p{3cm} p{3cm} p{6cm}}
$\psi=1$,  if $F_1<F_2$,
 $\psi=0$,  if $F_1>F_2$, and that $0\leq \psi\leq 1$ when $F_1=F_2$.
  %\end{tabular}
  
  If $\psi, F_1$ and $F_2$ are functions of some arguments, this agreement 
should be applied to their values calculated at any given arguments.

\begin{theorem}\label{Reduction.Th2}
 For all tests $(\psi, 
\phi)\in\mathscr S_0$   it holds
\begin{equation}\label{a7}
 L(\psi, \phi;\lambda_{0},\lambda_{1}) \geq \sum_{k=1}^{\infty} \sum_{n\in 
\mathscr G^k} p(n)E_{0} 
s^\psi_{n} \big(c({ n}) +g(z_{ n};\lambda_{0},\lambda_{1}) \big).
\end{equation}
There is an  equality in (\ref{a7})  if and only if  the following 
condition is satisfied:\\
{\bf Condition $\mathfrak D\bs{(\psi,\phi)}$}.
For all $k\geq 1$ and for all
$ n\in \mathcal{P}_{k}$
 \begin{equation}\label{Reduction.Eq10}
\phi_{ n} \simeq
I_{\left\{\lambda_{0}/\lambda_1 \preccurlyeq z_n\right\}}
 \quad P_0\mbox{-almost surely on} \quad S_{n}^{\psi}.
\end{equation}
  \end{theorem}

The proof can be carried out  along the lines of the proof of Theorem 2.2 
in \cite{novikov09}.

 Theorem \ref{Reduction.Th2} is the first step in the 
Lagrange minimization  we discussed earlier in 
this section. It reduces the problem of minimization of  $ L(\psi, 
\phi;\lambda_{0},\lambda_{1}) $ to that of minimization of 
\begin{equation}\label{a8}
  L(\psi; \lambda_{0}, \lambda_{1}) = 
\inf_\phi L(\psi, 
\phi; 
\lambda_{0}, \lambda_{1})=\sum_{k=1}^\infty \sum_{n\in\mathscr G^k}p(n)
E_{0} s^\psi_{n} \big( c({ n}) + g(z_{ n};\lambda_0,\lambda_1)\big)
 \end{equation}
over all stopping rules 
$\psi\in{\mathscr F_0}$.

Indeed, if we have a stopping rule $\psi$ minimizing 
$L(\psi;\lambda_0,\lambda_1)$ over all stopping rules in $\mathscr F_0$, 
then, combining the optimal $\psi$ with 
any decision rule $\phi$ satisfying $$\phi_{ n} \simeq
I_{\left\{\lambda_{0}/\lambda_1 \preccurlyeq z_n\right\}}\quad 
\mbox{for all }\;n\in\mathscr G^k\;\mbox{ and} \;k=1,2,\dots $$ (cf. 
\eqref{Reduction.Eq10}), we obtain
$$L(\psi,\phi;\lambda_0,\lambda_1)=L(\psi;\lambda_0,\lambda_1)\leq 
L(\psi^\prime;\lambda_0,\lambda_1)=L(\psi^\prime,\phi;\lambda_0,
\lambda_1)\leq L(\psi^\prime,\phi^\prime;\lambda_0,\lambda_1)
$$
Consequently,  \eqref{a6} is satisfied for all 
$(\psi^\prime,\phi^\prime)\in\mathscr S_0.$

 In such a way, starting from this moment, our focus will  be on the 
problem of minimizing $L(\psi;\lambda_0,\lambda_1)$  over 
all stopping 
rules $\psi$.

In the meanwhile, a useful consequence of Theorem 
\ref{Reduction.Th2} can  already be 
drawn for a particular (in fact, non-sequential) case, when the number 
of groups  $N$ is fixed in advance (and $s^\psi_{ n}\equiv 1$ for all $ 
n\in\mathscr G^N$). Combining the result of  Theorem \ref{Reduction.Th2}  
with Lemma \ref{a-l1} one immediately obtains an alternative proof of  
Theorem 2.1 
in \cite{Mukhopadhyay}.

\section{Optimal random group-sequential tests}\label{a20}

\subsection{Optimal stopping on a finite horizon}

For any $N\geq 1$, we define the class of {\em truncated} stopping rules 
$\mathscr F^N$ as
\begin{equation}\label{a38}
\mathscr F^N = \{\psi:( 1-\psi_{n_1})(1-\psi_{(n_1,n_2)})\dots(1-\psi_{n})
\equiv 0\; \mbox 
{for all} \;{ n}\in \mathscr
G^{N}\}.
\end{equation}

%It is obvious that $\mathscr F^N\subset \mathscr F^{N+1}$ for all $N\geq 
%1$.

 %It follows from Lemma \ref{And1-1} in the Appendix that $\mathscr 
%F^N\subset \mathscr F$ for all $N\geq 1$.

Let $\mathscr S^N$ be the set of all group-sequential  tests 
$(\psi,\phi)$ with 
$\psi\in \mathscr F^N$.

 For $\psi\in \mathscr F^N$, let us denote $L_N(\psi)=L_N(\psi;\lambda_0,\lambda_1)=L(\psi;\lambda_0,\lambda_1)$ (see (\ref{a8})).

In this section we characterize the structure of all stopping rules $\psi\in \mathscr F^N$
which minimize
$L_N(\psi)$
over all $\psi\in\mathscr F^N$.

It follows from (\ref{a38}) that
\begin{equation}\label{a38a}
L_N(\psi)=\sum_{k=1}^{N-1}\sum_{n\in\mathscr G^k}p(n)
E_{0} s^\psi_{n} \big( c({ n}) + g(z_{ n})\big)+\sum_{n\in\mathscr G^N}p(n)
E_{0} t^\psi_{n} \big( c({ n}) + g(z_{ n})\big).\end{equation}
The minimization of (\ref{a38a}) is an optimal stopping problem, so
the solution is through the following variant of the 
backward induction. 

Let the functions $V_k^N=V_k^N(z)$, $z\geq 0$, 
$k=1,2, \ldots, N$, be defined in the following  way:
starting from 
\begin{equation}\label{a9} V_N^N(z) \equiv g(z),\end{equation}  define 
recursively, for all $z\geq 0$,
\begin{equation}\label{Trunc.Eq4}
V_{k-1}^N(z) = \min \left\{ g( z ),  \;\bar{c}_{k} + \sum_{n\in \mathscr 
G} p_k(n)
E_{0}V_k^N(zz_{n})
\right\},\quad k=N, N-1, \dots,2.
\end{equation}
Let us denote
\begin{equation}\label{EVi}
\bar{V}_k^N(z)= \sum_{n\in\mathscr G}p_k(n)E_{0} V_k^N \left(z 
z_{n}\right),\quad z\geq 0,\quad k=1,2,\dots, N.
\end{equation}

It is important to bear in mind that all the functions above are 
constructed on the basis of the constants $\lambda_{0}>0$ and 
$\lambda_{1}>0$ and some 
definite cost structure ($c(m)$). Unfortunately, there is no satisfactory 
way to make  them all  explicit in the notation, so we leave  them 
implicit 
in all the elements of the construction.

The following theorem characterizes the structure of all  truncated 
stopping rules minimizing $L_N(\psi)$.

\begin{theorem}\label{Trunc.Th1}
For every $\psi\in \mathscr F^N$ 
\begin{equation}\label{Trunc.Eq6}
L_N(\psi) \geq \bar c_1 +\bar{V}_1^N(1).
\end{equation}
There is an equality in (\ref{Trunc.Eq6}) if  $\psi\in\mathscr F^N$  
satisfies the following \\
{\bf Condition $\bs{\mathfrak S_N(\psi)}$.}
 For all $\;1\leq k<N$ and all
${n} \in \mathcal{P}_{k}$ 
\begin{equation}\label{Trunc.Eq7}
 \psi_{ n}
\simeq
I_{\{g(z_{ n}) \preccurlyeq\bar c_{k+1} + \bar{V}_{k+1}^N( z_{ n})\} 
} \quad P_0\mbox{-almost surely on} \; T_{ n}^{\psi}.\end{equation}

Conversely, if there is an equality in (\ref{Trunc.Eq6}) for some 
$\psi\in\mathscr F^N$ then $\psi$ satisfies Condition $\mathfrak S_N(\psi)$.
\end{theorem}

The proof of Theorem \ref{Trunc.Th1} is placed in the Appendix.

\begin{corollary}\label{And1-2}
 Let  a truncated sequential test $(\psi,\phi)\in \mathscr S^N$  be such 
that
  Condition $\mathfrak S_N(\psi)$ of Theorem \ref{Trunc.Th1}
 and  Condition $\mathfrak D(\psi,\phi)$ of Theorem 
\ref{Reduction.Th2} are satisfied. 

Then for all truncated tests $(\psi^\prime,\phi^\prime)\in \mathscr S^N$ 
such that 
\begin{equation}\label{And1-3}
\alpha(\psi^{\prime},\phi^{\prime})\leq 
\alpha(\psi,\phi)\quad\mbox{and}\quad
\beta(\psi^{\prime},\phi^{\prime})\leq \beta(\psi,\phi)
\end{equation}
it holds
\begin{equation}\label{And1-4}
K_0(\psi^\prime)\geq K_0(\psi).
 \end{equation}
The inequality in (\ref{And1-4}) is strict if at least one of the 
inequalities in (\ref{And1-3}) is strict.

If there are equalities in all the inequalities in (\ref{And1-3}) and
(\ref{And1-4}), then 
 Condition $\mathfrak S_N(\psi^\prime)$ 
 and Condition $\mathfrak D(\psi^\prime,\phi^\prime)$  are satisfied for 
$(\psi^\prime,\phi^\prime)$. 
\end{corollary}

Corollary \ref{And1-2} is a direct consequence of Theorem 
\ref{Reduction.Th2} and Theorem \ref{Trunc.Th1} in combination with Lemma 
\ref{a-l1}.

\subsection{Optimal stopping on infinite horizon}\label{a21}

Very much like in \cite{novikov09}, 
the idea of this part is to pass to the limit, as $N\to\infty$, 
on both sides of (\ref{Trunc.Eq6}) in order to obtain a lower bound for 
the 
Lagrangian function and conditions to attain it.

Let us analyse first the right-hand side of (\ref{Trunc.Eq6}) supposing 
that $N\to\infty$.

We have:
\begin{equation}\label{And1-5}V_N^N(z)\geq V_N^{N+1}(z)          
         \end{equation}
         for all $z\geq 0$, because, by (\ref{Trunc.Eq4}), 
$$V_N^N(z)=g(z)\geq V_N^{N+1}(z)=
 \min \left\{ g( z ),  \;\bar{c}_{N+1} + \sum_{n\in 
\mathscr 
G} p_{N+1}(n)
E_{0}V_{N+1}^{N+1}(zz_{n})
\right\}.$$

Applying (\ref{Trunc.Eq4}) to (\ref{And1-5}) again, we obtain
$V_{N-1}^N(z)\geq V_{N-1}^{N+1}(z)$, and so on,  getting finally to 
\begin{equation}\label{And1-6}
V_k^N(z)\geq V_k^{N+1}(z),\quad z\geq 0, 
\end{equation}
 for any $k$ fixed.
It follows from (\ref{And1-6}) that there exists 
$\lim_{N\to\infty}V_k^N(z)=V_k(z)$, $z\geq 0$,
and, by the Lebesgue theorem of dominated convergence, 
$\lim_{N\to\infty}\bar V_k^N(z)=\bar V_k(z)$, $z\geq 0$.

 To pass to the limit on the left-hand side of (\ref{Trunc.Eq6}), 
let us define {\em truncation}  of any $\psi\in \mathscr F_0$ at any level $N$ as $\psi^N=(\psi_{n_1},\psi_{n_2}, \dots,\psi_{n_{N-1}},1,\dots)$ 
for all $n=(n_1,n_2,\dots,n_{N-1})\in \mathscr G^{N-1}$. Because $\psi^N\in 
\mathscr F^N$, we can apply
 (\ref{Trunc.Eq6})
with $L_N(\psi)=L(\psi^N)$. It is easy to see that $L_N(\psi)$ can be calculated directly over (\ref{a38a}), whatever be a stopping rule $\psi\in\mathscr F_0$.
\begin{lemma}\label{And1-7}
For any $\psi\in \mathscr F_0$ 
\begin{equation}\label{And1-8}
\lim_{N\to\infty} L_N(\psi)= L(\psi).
\end{equation} 
\end{lemma}
\begin{lemma}\label{And3-1}
 $$
  \inf_{\psi\in \mathscr F_0}L(\psi)=\bar c_1+ \bar V_1(1).
 $$
\end{lemma}
We place the proofs of Lemma \ref{And1-7} and Lemma \ref{And3-1} in the 
Appendix.
 
Now, we are able to characterize optimal stopping rules on infinite horizon.
\begin{theorem}\label{And1-9}
For every $\psi\in \mathscr F_0$ 
\begin{equation}\label{And1-10}
L(\psi) \geq \bar c_1 +\bar{V}_1(1).
\end{equation}
There is an equality in (\ref{And1-10}) if  $\psi\in\mathscr F_0$  
satisfies  

{\bf Condition $\bs{\mathfrak S_\infty(\psi)}$.}
 For all $\;1\leq k<\infty$ and all
${n} \in \mathcal{P}_{k}$ 
\begin{equation}\label{16-1029}
 \psi_{ n}
\simeq
I_{\{g(z_{ n}) \preccurlyeq\bar c_{k+1} + \bar{V}_{k+1}( z_{ n})\} 
}  \quad P_0\mbox{-almost surely on} \;T_{ n}^{\psi}.\end{equation}

Conversely, if there is an equality in (\ref{And1-10}) for some 
$\psi\in\mathscr F_0$ then $\psi$ satisfies Condition $\mathfrak S_\infty(\psi)$. 
\end{theorem}
The proof of Theorem \ref{And1-9} is laid out in the Appendix.

\begin{corollary}\label{And2-1}
 Let   a sequential test $(\psi,\phi)\in \mathscr S_0$  be such 
that
  Condition $\mathfrak S_\infty(\psi)$ of Theorem \ref{And1-9}
 and  Condition $\mathfrak D(\psi,\phi)$ of Theorem 
\ref{Reduction.Th2} are satisfied. 

Then for all sequential tests $(\psi^\prime,\phi^\prime)\in \mathscr S_0$ 
such that 
\begin{equation}\label{And2-3}
\alpha(\psi^{\prime},\phi^{\prime})\leq 
\alpha(\psi,\phi)\quad\mbox{and}\quad
\beta(\psi^{\prime},\phi^{\prime})\leq \beta(\psi,\phi)
\end{equation}
it holds
\begin{equation}\label{And2-4}
K_0(\psi^\prime)\geq K_0(\psi).
 \end{equation}
The inequality in (\ref{And2-4}) is strict if at least one of the 
inequalities in (\ref{And2-3}) is strict.

If there are equalities in all the inequalities in (\ref{And2-3}) and
(\ref{And2-4}), then 
 Condition $\mathfrak S_\infty(\psi^\prime)$ 
 and Condition $\mathfrak D(\psi^\prime,\phi^\prime)$  are satisfied for 
$(\psi^\prime,\phi^\prime)$. 
\end{corollary}

\section{Optimality of the random sequential probability ratio test}\label{sec5}

In this section we apply the general results of preceding sections to 
a particularly important model, assuming that the groups for the 
group-sequential test are formed in a stationary way. More precisely, we 
assume in this section that the group  sizes $\nu_1,\nu_2,\dots$ are 
identically distributed (and their common distribution 
is given by $p(n)=P(\nu_k=n)$, for $n\in \mathscr 
G$ and $k=1,2,\dots$). 
Respectively, the average group costs $\bar c_i=\bar c$ keep the same
value over the experiment time.
 We will characterize the structure of optimal group-sequential tests in 
 the case of infinite horizon. In particular, we will  prove the optimal 
property of the random sequential 
probability ratio test (RSPRT) proposed by \cite{Mukhopadhyay} for the 
random group-sequential model.

%Let us analyze the structure of optimal group-sequential tests, assuming that 
%\begin{equation}\label{08-0904}P_0(f_1(X)>0)=1.\end{equation}

There are three constants involved in the construction of optimal tests in 
this case:
$\bar c$, $\lambda_0$ and $\lambda_1$. It is easy to see that only two of 
them suffice  to obtain  all the optimal 
sequential tests of Corollary \ref{And2-1}. Let these be $c=\bar c>0$ 
and $\lambda=\lambda_0>0$ (and just assume that   $\lambda_1=1$). The 
structure of 
optimal tests of Theorem \ref{And2-1}  now
acquires  a simpler form. %As before, the notation we use does 
%not have explicit reference to them, unless otherwise  stated.  

    Let 
    \begin{equation}\label{05-2005}\rho_0(z)=\rho_0(z;c, \lambda)=g(z;\lambda)=\min\{\lambda, z\},\quad z\geq 0, \end{equation} 
    and, 
recursively over 
$k=1,2,\dots$, 
\begin{equation}\label{08-0941}
\rho_k(z)=\rho_k(z;c,\lambda)=\min\{g(z;\lambda), c+\sum_{n\in \mathscr G}p(n)E_0\rho_{k-1}(z z_n;c,\lambda) 
\}, \; z\geq 0.
\end{equation}
Let also
\begin{equation}\label{08-0959}
\bar \rho_k(z)=\bar\rho_k(z;c,\lambda)=\sum_{n\in \mathscr G}p(n)E_0\rho_{k}(z z_n;c,\lambda). 
\end{equation}

It follows from \eqref{a9} - \eqref{Trunc.Eq4}
that
\begin{equation}\label{08-1013}
V_k^N(z)=\rho_{N-k}(z;c,\lambda), \; z\geq 0 ,
\end{equation}
and from  \eqref{EVi},
\begin{equation}\label{08-1017}
 \bar V_k^N(z)=\bar \rho_{N-k}(z;c,\lambda).
\end{equation}

Let us define 
\begin{equation}\label{06-0950}\rho(z)=\rho(z;c,\lambda)=\lim_{k\to\infty}\rho_k(z;c,\lambda),\quad z\geq 0.\end{equation}

If we take the limit, as ${N\to\infty}$, in \eqref{08-1013}, then
\begin{equation}\label{08-1024}
V_k(z)=\rho(z;c,\lambda), \; z\geq 0,
\end{equation}
for all $k=1,2,\dots$, and by \eqref{08-1017}
\begin{equation}\label{08-1115}
\bar V_k(z)=\bar \rho(z;c,\lambda), \; z\geq 0.
\end{equation}
Stopping rule \eqref{16-1029} in Condition $\mathfrak S_\infty(\psi)$  now 
transforms
to
 \begin{equation}\label{15-1917}
 \psi_{n}
\simeq
I_{\{g(z_{ n};\lambda) \preccurlyeq c + \bar{\rho}( z_{ n};c,\lambda)\}}, 
\end{equation}
so the form of optimal stopping rules entirely depends on whether the 
inequality
\begin{equation}\label{08-1128}
 g(z;\lambda) \leq  c + \bar{\rho}( z;c,\lambda)
\end{equation}
(and/or its strict variant) is fulfilled or not at $z=z_n$.

First of all, it is easy to see that if 
\begin{equation}{\label{25-1908}}\lambda<c+  \bar{\rho}( \lambda;c,\lambda)\end{equation}
then \eqref{15-1917} implies that $\psi_n\equiv 1$ for all $n\in \mathscr G$ (the optimal test stops 
after the first group is taken). Therefore, non-trivial  optimal sequential tests are only obtained  if
\begin{equation}\label{22-161an}\lambda>c+  \bar{\rho}( \lambda;c,\lambda), \end{equation}
which will be assumed in what follows.
\begin{lemma}\label{22-161} If
\begin{equation}\label{18-1107}
 P_{0}(f_{1}(X)>0)=1,
%\mbox{and}\quad P_{\theta_1}(f_{\theta_0}(X)>0)=1. 
\end{equation}
then for any positive $c$ and $\lambda$ satisfying \eqref{22-161an}
 there exist $0<A<\lambda$ and $B>\lambda$, such that
\begin{equation}\label{22-1619}
 g(A;\lambda)=c+  \bar{\rho}(A;c,\lambda), \mbox{}\quad g(B;\lambda)=c+  \bar{\rho}(B;c,\lambda),
\end{equation}
and
 \begin{equation}\label{22-1632}
 g(z;\lambda)<c+  \bar{\rho}(z;c,\lambda)\quad  \mbox{for all}\quad  0\leq z<A\quad
 \mbox{and all}\quad z>B,
\end{equation}
 and
\begin{equation}\label{22-1634}
 g(z;\lambda)>c+  \bar{\rho}(z;c,\lambda)\quad \mbox{ for all}\quad  A<z< B.
\end{equation}
\end{lemma}

%\begin{textsc}{Theorem \ref{t6}}
% {Proof of Theorem \ref{t6}}  
It follows from Lemma \ref{22-161} that, if  \eqref{18-1107} holds then 
  (\ref{15-1917}) is equivalent to
  \begin{equation}\label{22-1659}
     I_{\{z_n\in (A,B) \}} \leq 1-\psi_n\leq I_{\{z_n\in [A,B]\}},
  \end{equation}
that is, any optimal test is a randomized version of  the random sequential probability ratio test (RSPRT) 
 by \cite{Mukhopadhyay} which, in our terms, can be described  as $(\psi,\phi)$ 
with 
\begin{equation}\label{30-1702}
\psi_n =I_{\{z_n\not\in (A,B) \}}\quad\mbox{and}\quad\phi_n=I_{\{z_n\geq B\}},
\end{equation}
for all $n\in\mathscr G^k$ and $k\geq 1$.

Obviously, $\psi_n$ in \eqref{30-1702} is a particular case of \eqref{22-1659}, and $\phi_n$  satisfies 
Condition $\mathfrak D(\psi,\phi)$. In addition, by virtue of  Theorem 3.1 in \cite{Mukhopadhyay}, 
$(\psi,\phi)\in \mathscr S_0$ 
(the details can be found in the proof of Theorem \ref{t6} below). 
 Consequently, it  follows  that this RSPRT  is optimal in the sense of 
Corollary \ref{And2-1}.

In the same way, all the sequential tests $(\psi,\phi)$ with $\psi$ satisfying \eqref{22-1659}
and $\phi$ satisfying \eqref{30-1702} (for all $n\in\mathscr G^k$ and $k\geq 1$) share the optimum property with the RSPRT,
when 
\eqref{18-1107} is satisfied. In particular, obviously,  this is the case when  the hypothesized distributions belong to a Koopman-Darmois family.

If \eqref{18-1107} is not satisfied,  optimal tests with stopping rules (\ref{15-1917}) are not necessarily of the RSPRT type.
This can be seen from the following simple example. 

Let $H_0$ state that the (one per group) observations  follow a uniform distribution on [0,1], whereas 
under $H_1$ they are
assumed  to be uniform on $[0,0.5]$. Using definition of $\rho$ in \eqref{06-0950}, on the basis of $\rho_k$ defined in
\eqref{05-2005} and \eqref{08-0941}, with $\lambda=2$ and $c=1$,
one easily sees that $\rho(z;c,\lambda)=\rho(z;1,2)=g(z;2)=\min\{z,2\}$ and $c+\bar \rho(z;c,\lambda)=1+\bar\rho(z;1,2)=
1+\min\{z,1\}$. Let us consider an (optimal) test corresponding to  \eqref{08-1128}, with a strict inequality. It it immediate that, with the above definitions of $g$ and $\bar\rho$, \eqref{08-1128} holds if and only if $z< 2$. But the consecutive values of 
the probability ratios
$z_1,z_2,\dots, z_k $ are, respectively, $2,4,8,\dots, 2^k$, whenever $X_1,X_2,\dots, X_k \leq 0.5$, so the test (minimizing $K_0$)  only stops when, for the first time, 
$X_i>0.5$, in which case  $z_i=0$.

It is seen from this example, first, that the test minimizing $K_0$ is not an RSPRT (because an RSPRT should also stop when $z_i\geq B>0$ which happens, under $H_0$, with a positive probability, thus, there would be a positive $\alpha$-error), and  second, that,  in no way, it
minimizes $K_1$, because under $H_1$ it never stops.

Nevertheless, the following theorem shows  that, even if \eqref{18-1107} is not satisfied, not only the  RSPR tests with $\psi$ satisfying \eqref{30-1702}, 
but also their
``randomized" versions with $\psi$ satisfying \eqref{22-1659}, are optimal in the sense of \cite{waldwolfowitz}, i.e. they 
minimize the average cost under both 
$H_0$ and $H_1$, given  restrictions on the error probabilities. 

\begin{theorem}\label{t6}
Let $A<B$  be two positive  constants.
Let $\psi$ be any stopping rule satisfying \begin{equation}\label{13-0929}
I_{\{  z_n\in(A,B) \}}\leq 1-\psi_n\leq I_{\{  z_n\in 
[A,B] \}},
    \end{equation}
for all $n\in \mathscr G^k$ and $k=1,2,\dots$, and let $\phi$ be a 
decision rule defined as
\begin{equation}\label{13-1011}               
\phi_n = I_{\{z_n\geq B\}}
\end{equation}
for all $n\in \mathscr G^k$ and $k=1,2,\dots$.

%Condition $\mathfrak D(\phi)$ of Theorem \ref{Reduction.Th2} satisfied. 
Then $(\psi,\phi) \in \mathscr S_0\cap\mathscr S_1$,
and it  is optimal 
 in
the following sense: for any sequential test $(\psi^\prime,\phi^\prime)\in 
\mathscr S_0\cap\mathscr S_1$ such that
\begin{equation}\label{93}
\alpha(\psi^\prime,\phi^\prime)\leq\alpha(\psi,\phi)\quad
\mbox{and}\quad\beta(\psi^\prime,\phi^\prime)\leq\beta(\psi,\phi)
\end{equation}
it holds
\begin{equation}\label{94}
K_0(\psi)\leq K_0 (\psi^\prime)\quad 
\mbox{and}\quad
K_1(\psi)\leq K_1 (\psi^\prime).
\end{equation}
Both inequalities in (\ref{94}) are strict if at least one of the
inequalities in (\ref{93}) is strict. 
\end{theorem}

\begin{remark} \sl
The optimum property stated in Theorem \ref{t6}, in the case of one-per-group observations is known as the Wald-Wolfowitz optimality 
(see \cite{waldwolfowitz}). \cite{Burkholder} proved that all ``extended" SPRTs, i.e. those admitting a
 randomized decision between stopping and continuing in case $z_n=A$ or $z_n=B$, share the same optimum property with the 
 SPRT. Our Theorem \ref{t6} states the same, in the case of random group sequential tests: the ``extended" group-sequential
 tests, i.e. those with stopping rules satisfying \eqref{13-0929}, minimize the average cost under both $H_0$ and $H_1$.
 \end{remark}
\begin{remark} \sl
Very much like in the classical one-per-group case, when taking no observations is permitted, only the 
case $1 \in [A,B]$ is meaningful for the RSPRT, because otherwise a trivial 
test (namely the one which, without  taking any observations, accepts or rejects $H_0$ depending on whether $1<A$ or $1> B$) 
performs better than the optimal tests of Theorem 
\ref{t6} ($\min\{1,\lambda\}<c+\bar\rho(1;c,\lambda)$ in terms of optimal stopping of Theorem \ref{And1-9}). 
\end{remark}

\section*{Acknowledgment}
A. Novikov thanks SNI by CONACyT, Mexico,  for a partial 
support for this work. X.I. Popoca-Jiménez thanks CONACyT, Mexico, for 
scholarships for her studies.

\appendix
\section{Appendix}

In this section, lengthy or too  technical proofs are gathered together. 

For simplicity, it is assumed throughout this section that all  group sizes are identically distributed, even when
this is not explicitly required  in the respective statement.    

\begin{lemma} \label{06-1343}
 If a stopping rule $\psi$ is such that 
\begin{equation}\label{06-1343a}
 \sum_{m\in\mathscr 
G^r}p(m)E_{\theta}{t_{n,m}^\psi}\to 0\,\mbox{ as}\; r\to\infty,
\end{equation}
for some $n\in \mathscr G^k$, $k\geq 1$, then
\begin{equation}\label{06-1343b}
E_{\theta} s_n^{\psi}+\sum_{r=1}^\infty\sum_{m\in 
\mathscr G^r}p(m)E_{\theta} s_{n,m}^{\psi} = E_{\theta}t_n^\psi.
\end{equation}
\end{lemma}

{\bf Proof of Lemma \ref{06-1343}}
  Let $r$ be any natural number. Then
\begin{eqnarray*}
&&\sum_{m\in \mathscr G^r}p(m)E_{\theta} t_{n,m}^\psi -\sum_{m\in \mathscr G^{r+1}}p(m)E_{\theta}t_{n,m}^\psi = \nonumber \\
&=&\sum_{m\in \mathscr G^r}p(m)E_{\theta}t_{n,m}^\psi -\sum_{m\in \mathscr G^{r}}\sum_{i\in \mathscr G}
p(m)p(i)E_{\theta}t_{n,m,i}^\psi \nonumber\\
&=&\sum_{m\in \mathscr G^r}p(m)\big(E_{\theta} t_{n,m}^\psi-\sum_{i\in \mathscr G}p(i)E_{\theta}t_{n,m,i}^\psi\big) = \nonumber \\
&=&\sum_{m\in \mathscr G^r}p(m)\big(E_{\theta}t_{n,m}^\psi-E_{\theta}(1-\psi_{n_1})\dots(1-\psi_n)(1-\psi_{n,m_1})\dots(1-\psi_{n,m})\big) \nonumber \\
&=&\sum_{m\in \mathscr G^r}p(m)E_{\theta}s_{n,m}^\psi,
\end{eqnarray*}	
so
\begin{equation}\label{06-1343c}
\sum_{m\in \mathscr G^r}p(m)E_{\theta}s_{n,m}^\psi= \sum_{m\in \mathscr G^r}p(m)E_{\theta}t_{n,m}^\psi 
-\sum_{m\in \mathscr G^{r+1}}p(m)E_{\theta}t_{n,m}^\psi. 
\end{equation}
Applying the sum over $r$ from  $r=1$ to $r=k$ on both sides of \eqref{06-1343c},  we obtain
\begin{eqnarray}\label{06-1343d}
&&\sum_{r=1}^k \sum_{m\in \mathscr G^r}p(m)E_{\theta}s_{n,m}^\psi =\sum_{r=1}^k \big( \sum_{m\in \mathscr G^r}p(m)E_{\theta}t_{n,m}^\psi 
-\sum_{m\in \mathscr G^{r+1}}p(m)E_{\theta}t_{n,m}^\psi \big)= \nonumber\\ \label{29-2025}
&=&  \sum_{m\in \mathscr G}p(m)E_{\theta} t_{n,m}^{\psi}-\sum_{m\in \mathscr{G}^{k+1}} p(m) E_{\theta}t_{n, m}^{\psi}. 
\end{eqnarray}
Passing to the limit, as $k\to\infty$, in \eqref{06-1343d} and making use of \eqref{06-1343a} on the right-hand side,  we get
\begin{equation}\label{06-1343f}
\sum_{r=1}^\infty \sum_{m\in \mathscr G^r}p(m)E_{\theta}s_{n,m}^\psi =  \sum_{m\in \mathscr G}p(m)E_{\theta} t_{n,m}^{\psi}. 
\end{equation}

On the other hand, it is easy to see (by definitions \eqref{31-1252}-\eqref{31-1253}) that
\begin{equation}\label{30-1036}
E_{\theta}s_{n}^\psi= E_{\theta}t_{n}^\psi- \sum_{m\in \mathscr G}p(m)E_{\theta}t_{n,m}^\psi.                    
                   \end{equation}
Adding   the expressions on both sides of  \eqref{30-1036}  to the respective sides of \eqref{06-1343f}              
  we obtain \eqref{06-1343b}.  $\Box$ 

\begin{lemma}\label{And1-1}
For a stopping rule $\psi$, it holds 
 \begin{equation}\label{And1-1a}
\sum_{m\in\mathscr 
G^r}p(m)E_\theta{t_{m}^\psi}\to 0,\,\mbox{ as}\; r\to\infty,
\end{equation}
if and only if
\begin{equation}\label{And1-1b}
\sum_{r=1}^\infty\sum_{m\in 
\mathscr G^r}p(m)E_\theta s_{m}^{\psi} = 1.
\end{equation}
\end{lemma}

{\bf Proof of Lemma \ref{And1-1}}
Let us suppose that \eqref{And1-1a} is satisfied.
Repeating the steps of the proof of Lemma \ref{06-1343}, but with $t_{m}^\psi$ instead of $t_{n,m}^\psi$, we obtain,
in place of \eqref{06-1343f},
\begin{equation}\label{07-0349}
\sum_{r=1}^\infty \sum_{m\in \mathscr G^r}p(m)E_{\theta}s_{m}^\psi =  \sum_{m\in \mathscr G}p(m)E_{\theta} t_{m}^{\psi}=1,
\end{equation}
the latter because $ t_{m}^{\psi}=1$ for all $m\in\mathscr G$, by definition.
\eqref{And1-1b} is proved.

Let now \eqref{And1-1b} be fulfilled.
Acting as in the proof of Lemma \ref{06-1343} again, we obtain
\begin{eqnarray}\label{29-2009}
&&\sum_{r=1}^k \sum_{m\in \mathscr G^r}p(m)E_{\theta}s_{m}^\psi =
 \sum_{m\in \mathscr G}p(m)E_{\theta} t_{m}^{\psi}-\sum_{m\in \mathscr{G}^{k+1}} p(m) E_{\theta}t_{ m}^{\psi}. 
\end{eqnarray}
The  left-hand side of \eqref{29-2009} tends to 1, as $k\to\infty$, by virtue of \eqref{And1-1b}. The first term on the right-hand size
is equal to 1, for the same reason as in \eqref{07-0349}. Therefore, the second term on the right-hand side of \eqref{29-2009} tends to 0, so
\eqref{And1-1a} follows.$\Box$

\begin{lemma}\label{06-1115}
 Let $\psi\in \mathscr F_0$ be such that $K_0(\psi)<\infty$.
 Then
 \begin{equation}\label{06-1116}
  \sum_{ n\in\mathscr G^{N}} p( n)  c( n)E_{0} 
t_{ n}^\psi \to 0,\; \mbox{as}\; N\to\infty.
 \end{equation}
\end{lemma}
{\bf Proof of Lemma \ref{06-1115}} Let $\psi\in \mathscr F_0$ be such that $K_0(\psi)<\infty$.
By definition,
$$
K_0(\psi)= \sum_{k=1}^{\infty}\sum_{ n\in\mathscr G^{k}} p( n)  c( n)E_{0} 
s_{ n}^\psi,
$$
thus,
\begin{equation}\label{06-1203}
\sum_{k=N}^{\infty}\sum_{ n\in\mathscr G^{k}} p( n)  c( n)E_{0} 
s_{ n}^\psi\to 0, 
\end{equation}
as $N\to\infty$. Let $c(n|N)=c(n_1)+\dots+ c(n_N)$ for any
$n\in\mathscr 
G^k$ and $k\geq N$.

It follows from Lemma \ref{06-1343}  that
\begin{equation}
\sum_{n\in \mathscr G^N}c(n)p(n)E_0t_n^\psi
=\sum_{n\in \mathscr G^N}c(n)p(n)
(E_0s_n^\psi +\sum_{r=1}^\infty\sum_{m\in \mathscr G^r} p(m) 
E_0s_{n,m}^\psi)
\end{equation}

\begin{equation}
=\sum_{k=N}^{\infty} \sum_{ n\in\mathscr G^{k}}c( 
n|N) p( n) E_{0} 
s_{ n}^\psi\leq \sum_{k=N}^{\infty}\sum_{ n\in\mathscr G^{k}} p( n)  c( 
n)E_{0} 
s_{ n}^\psi\to 0,
\end{equation}
as $N\to\infty$, by virtue of \eqref{06-1203}.
$\Box$

\subsection*{Proof of Theorem \ref{Trunc.Th1}}

Let $\psi\in\mathscr F^N$ be any truncated stopping rule. For any $k=1, 2, 
\ldots, N$, let us define 
\begin{eqnarray}
Q_k^N(z)&=& \sum_{i=1}^{k-1} { \sum_{n\in\mathscr G^i } }p(n)
E_{0}
s^\psi_{ n} \big( c(n) + g(z_{n}) \big)\nonumber\\
\label{And1}
&+& {\sum_{n\in \mathscr G^k}}p(n) E_{0} t_{n}^\psi \big( c( n) + 
V_k^N(z z_n)\big),
\end{eqnarray}%
for all $z\geq 0$.
In particular, we have $Q_N^N(1)=L_N(\psi)$ and $Q_1^N(1)=\bar c_1+\bar 
V_1^N(1)$.
 
We want to prove
\begin{equation}\label{Trunc.Apx.Eq2}
L_N(\psi)=Q_N^N(1)\geq Q_{N-1}^N(1)\geq \dots\geq Q_1^N(1)=\bar c_1+\bar 
V_1^N(1)
\end{equation}
and determine how should be $\psi\in \mathscr F^N$ in order to turn all the 
inequalities in (\ref{Trunc.Apx.Eq2}) into equalities.

Let us first prove that for
$1\leq k\leq N-1$ it holds
\begin{equation}\label{Trunc.Apx.EqQ}
Q_{k+1}^{N}(1) \geq Q_{k}^{N}(1).
\end{equation}

By definition (\ref{And1}), inequality \eqref{Trunc.Apx.EqQ} is equivalent 
to
\begin{eqnarray}\label{Trunc.Apx.Eq3}
\sum_{ n\in\mathscr G^k} p({ n}) E_{0} s_{ n}^\psi \big( c({ n}) 
+ g(z_{ n}) \big) &+&  \sum_{ n\in\mathscr G^{k+1}} p({ n}) E_{0} 
t_{ n}^\psi \big( c({ n}) + V_{k+1}^N(z_{ n})  \big)
\nonumber\\
&\geq& \sum_{{ n}\in\mathscr G^k} p({ n}) E_{0} t_{ n}^\psi \big( 
c({ n}) + V_{k}^N( z_{ n}) \big).
\end{eqnarray}

By 
the Fubini 
theorem, the left-hand side of \eqref{Trunc.Apx.Eq3} is equal to 
\begin{multline}\label{Trunc.Fubini}
\sum_{ n\in\mathscr G^k} p( n) \int t_{ n}^\psi \Big( c( n)+  \psi_{ n} 
g(z_ n)   
 +( 1 - \psi_{ n} )\big( \sum_{m\in \mathscr G} p_{k+1}(m) 
c(m) 
\\
+ \sum_{m\in\mathscr G} p_{k+1}(m) 
\int V_{k+1}^N
( z_{ n} z_{m} ) f_{0}^{(m)} d\mu^{m}	 \big) \Big)
 f_{0}^{ (n)} d\mu^{ n}. 
\end{multline}
By virtue of Lemma 5.1 in \cite{novikov09} the minimum value of
(\ref{Trunc.Fubini}), over all $\psi_n$, $n\in \mathscr G^k$,  is 
equal to 
\begin{multline}\label{And2}
  \sum_{{n}\in\mathscr G^k} p({ n}) \int t_{ n}^\psi \Big( c({ 
n})
+ \min \Big\{ g(z_{ n}), 
 \sum_{m\in \mathscr G} p_{k+1}(m) 
c(m) 
 \\
 + \sum_{m\in\mathscr G} p_{k+1}(m) \int V_{k+1}^N( z_{ n} 
z_{m} )
f_{0}^{(m)} d\mu^{m}
\Big\} \Big) f_{0}^{( n)} d\mu^{ n}\\
= \sum_{{n}\in\mathscr G^k} p({ n}) E_{0} t_{ n}^\psi \big( c({ 
n}) + V_{k}^N( z_{ n}) \big),
\end{multline}
 and is attained 
 if and only if  for all ${ n}\in 
\mathcal{P}_{k} $ 
\begin{equation}\label{And3}
\psi_{ n}
\simeq
I_{\{g(z_{ n}) \preccurlyeq\bar c_{k+1} + \bar{V}_{k+1}^N( z_{ n})\} 
}\; \end{equation}
 $
  \mu^{ n}$-a.e. on $T_{ n}^{\psi}\cap \{f_0^{( n)}>0\}.
 $
This completes the proof of \eqref{Trunc.Apx.EqQ}.

Applying \eqref{Trunc.Apx.EqQ} consecutively for $k=1,2,\dots, N-1$ we also obtain the proof for \eqref{Trunc.Apx.Eq2}.
  In 
addition, there are equalities in all the inequalities in
\eqref{Trunc.Apx.Eq2} if and only if \eqref{And3} is satisfied for all ${ 
n}\in 
\mathcal{P}_{k} $
 $
  \mu^{ n}$-a.e. on $T_{ n}^{\psi}\cap \{f_0^{ (n)}>0\}
 $, for all $k=1,2,\dots,N-1$, which coincides with Condition $\mathfrak 
S_N(\psi)$ of Theorem \ref{Trunc.Th1}.

\subsection*{Proof of Lemma \ref{And1-7}}
 Let us suppose first  that $L(\psi)< \infty$.   \eqref{a8} and \eqref{a38a} we have
\begin{eqnarray}\label{And1-71}
 L(\psi)-L_{N}(\psi)
&= & \sum_{k=N}^{\infty} \sum_{n \in \mathscr G^{k}}  p(n) E_0 s_ n^{\psi}\; \big(c(n) +g (z_n) \big)\nonumber\\
& -&  \sum_{n \in \mathcal G^{N}} p(n)c(n) E_0t_{n}^{\psi} - \sum_{n \in \mathcal G^{N}} p(n)E_0 t_{n}^{\psi}g(z_n). 
\end{eqnarray}

Because the series  $L(\psi)$ is converging, by supposition, the first term on the right-hand side of \eqref{And1-71} 
tends to zero, as $N \to \infty$. 

In addition, this implies that $K_0(\psi)<\infty$, so by Lemma \ref{06-1115}  (see \eqref{06-1116}), 
the second term on the right-hand side of  \eqref{And1-71} also tends to zero as  $N \to \infty$. 

At last, $g(z)\leq \lambda_{0}$ for all non-negative $z$. Thus,
\begin{equation*}
\sum_{n \in \mathcal G^{N}} p(n)E_0 t_{n}^{\psi}g( z_n) \leq 
\lambda_{0}\sum_{n \in \mathcal G^{N}} p(n)E_0 t_{n}^{\psi} \to 0, \quad \text{as } N \to \infty.
\end{equation*}
This latter   holds by  Lema \ref{And1-1}, because the supposition $\psi \in \mathscr{F}$ implies \eqref{And1-1b}. 
Thus, we proved \eqref{And1-8}  
in case $L(\psi)< \infty$.

Let now $L(\psi)=\infty$, then 
\begin{equation*}
L_N(\psi) \geq \sum_{k=1}^{N-1} \sum_{n \in \mathcal G^{k}}  p(n) E_0 s_ n^{\psi}\; \big(c(n) +g (z_n) \big) 
\to L(\psi)= \infty, \quad \text{as } N\to \infty,
\end{equation*}
so that  \eqref{And1-8} holds in this case as well.  $\Box$

\subsection*{Proof of Lemma \ref{And3-1}} 
Denote
$U = \inf_{\psi\in\mathscr  F } L(\psi)$,
$U_N=\inf_{\psi\in\mathscr F^N} L(\psi)$.

By Theorem \ref{Trunc.Th1},
$ U_N = \bar c_1 + \bar V_1^N(1) $
for any $N=1,2,\dots$.
It is obvious that $U_N\geq U$ for any $N=1,2,\dots$, so,
$ \lim_{N\to\infty}U_N\geq U$.
Let us show that in fact
$\lim_{N\to\infty}U_N = U$.

Suppose the contrary:
$\lim_{N\to\infty} U_N=U+4\varepsilon$
with some
$\varepsilon>0$;
this would imply
\begin{equation}\label{And6}
U_N\geq U+3\varepsilon 
\end{equation}
for all large enough $N$.
By definition of $U$, there exists a stopping rule $\psi\in \mathscr F_0$, 
such that
$U\leq L(\psi)\leq U+\varepsilon$.
Since, by Lemma \ref{And1-7}, 
$L_N(\psi)\to L(\psi)$, as $N\to\infty$, we
have
$ L_N(\psi)\leq U+2\varepsilon $
for all large enough $N$. Because, by definition,
$U_N\leq L_N(\psi)$, we have that
$ U_N\leq U+2\varepsilon $
for all large enough $N$, which contradicts (\ref{And6}).

Hence,
$U = \lim_{N\to\infty} U_N
=  \bar c_{1} + \lim_{N\to\infty} \bar V_1^N(1)
=  \bar c_{1} + \bar V_1(1)$.
$\; \Box$

\subsection{Proof of Theorem \ref{And1-9}}
%\begin{lemma}\label{06-1122} $$\sum_{k=1}^\infty\sum_{n\in \mathscr 
%G^k}p(n)E_\theta s_n^\psi=1
%$$ if and only if 
%\begin{equation}\label{06-1123}
%  \sum_{ n\in\mathscr G^{k}} p( n) E_{\theta} 
%t_{ n}^\psi \to 0,\; \mbox{as}\; k\to\infty.
%\end{equation}
%\end{lemma}

Let $\psi\in\mathscr F_0$ be any stopping rule. For $k=1, 2, 
\ldots$, let us define for all $z\geq 0$

\begin{eqnarray}\label{And5-220}
Q_k(z)&=& \sum_{i=1}^{k-1} { \sum_{n\in\mathscr G^i } }p(n)
E_{0}
s^\psi_{ n} \big( c(n) + g(z_{n}) \big)\nonumber\\
&+& {\sum_{n\in \mathscr G^k}}p(n) E_{0} t_{n}^\psi \big( c( n) + 
V_k(z z_n)\big)
\end{eqnarray}%
 (cf. (\ref{And1})).  By the Lebesgue  theorem of dominated convergence, it 
follows that 
\begin{equation}\label{And5-243}
\lim_{N\to\infty}Q_k^N(z)=Q_k(z).
\end{equation}

For the same reason, we have from  (\ref{Trunc.Apx.EqQ})
\begin{equation}\label{07-0944}
Q_{k+1}(1) \geq Q_{k}(1).
\end{equation}
for any $k=1,2,\dots$ 

Completely analogously to obtaining \eqref{And3} from 
\eqref{Trunc.Apx.EqQ},
through the steps of (\ref{Trunc.Apx.Eq3}), 
(\ref{Trunc.Fubini}) and (\ref{And2}), we can obtain  that there is  
an equality in 
(\ref{07-0944}) if 
  for all ${ n}\in 
\mathcal{P}_{k} $ 
\begin{equation}\label{And35-1838}
\psi_{ n}
\simeq
I_{\{g(z_{ n}) \preccurlyeq\bar c_{k+1} + \bar{V}_{k+1}( z_{ n})\} 
}\; \end{equation}
 $
  \mu^{ n}$-a.e. on $T_{ n}^{\psi}\cap \{f_0^{( n)}>0\}.
 $
Applying (\ref{07-0944}) consecutively  for $k=1,2,\dots$, we obtain  
%(\ref{Trunc.Apx.Eq2})
\begin{equation}\label{07-0952}
Q_{k}(1)\geq Q_{k-1}(1)\geq\dots\geq Q_1(1)=\bar c_1+\bar 
V_1(1),
\end{equation}
and there are  all equalities in (\ref{07-0952}), for any natural $k$,
in case Condition $\mathfrak S_\infty(\psi)$ is satisfied.

In particular, we have that for all natural $k$
\begin{eqnarray}\nonumber
Q_{k}(1)&=&\sum_{i=1}^{k-1} { \sum_{n\in\mathscr G^i } }p(n)
E_{0}
s^\psi_{ n} \big( c(n) + g(z_{n}) \big)\\ \label{07-1155}
&+& {\sum_{n\in \mathscr G^k}}p(n) E_{0} t_{n}^\psi \big( c( n) + 
V_k(z_n)\big)=\bar c_1+\bar 
V_1(1).
\end{eqnarray}
It follows from this, first, that $K_0(\psi)<\infty$, because otherwise
$$\lim_{k\to\infty}\sum_{i=1}^{k-1} { \sum_{n\in\mathscr G^i } }p(n)
E_{0}
s^\psi_{ n} \big( c(n) + g(z_{n}) \big)\geq K_0(\psi)
$$
would be infinite, and hence, would contradict (\ref{07-1155}).
Thus, $K_0(\psi)<\infty$, so, 
by virtue of Lemma \ref{06-1115},
\begin{equation}\label{07-1214}
\sum_{n\in \mathscr G^k}p(n)c( n) E_{0} t_{n}^\psi \to 0, \;\mbox{as}\; 
k\to\infty.
\end{equation}
Furthemore,
\begin{equation}\label{07-1215}
\sum_{n\in \mathscr G^k}p(n) E_{0} t_{n}^\psi  
V_k(z_n)\leq \lambda_0 \sum_{n\in \mathscr G^k}p(n) E_{0} t_{n}^\psi \to 0, 
\;\mbox{as}\; 
k\to\infty,
\end{equation}
because $\psi\in \mathscr F_0$ by supposition.

It follows from (\ref{07-1155}), (\ref{07-1214}) and (\ref{07-1215}) that
$$\lim_{k\to\infty }\sum_{i=1}^{k-1} { \sum_{n\in\mathscr G^i } }p(n)
E_{0}
s^\psi_{ n} \big( c(n) + g(z_{n}) \big)=L(\psi)=\bar c_1+\bar 
V_1(1),
$$
thus, the sufficiency part of Theorem \ref{And1-9} is proved. 

Let us prove the necessity part.

Passing to the limit in \eqref{Trunc.Apx.Eq2} as $N\to\infty$ we have for any $\psi\in \mathscr F_0$
\begin{equation}\label{07-1237}
L(\psi)\geq 
Q_{k}(1)\geq Q_{k-1}(1)\geq\dots\geq Q_1(1)=\bar c_1+\bar 
V_1(1),
\end{equation}
for all natural $k$. 
Let us suppose now that there is an equality in \eqref{And1-10} for a 
stopping rule $\psi\in\mathscr F_0$. 
It follows from \eqref{07-1237} now that all the inequalities in 
\eqref{07-1237} are in fact equalities, so $Q_{k+1}(1)= Q_k(1)$ for all 
natural $k$. This implies,
once again, that $\mathfrak S_\infty(\psi)$ is satisfied. $\Box$
\subsection{Proof of Theorem \ref{t6}}
Before starting with the proof of the Theorem, we need some previous work to be done.

Let us denote 
\begin{equation}\label{17-1048}
R(z;c,\lambda)=\inf_{(\psi,\phi)\in \mathscr{S}_0}
(cE_0\tau_\psi+\lambda\alpha(\psi,\phi)+z\beta(\psi,\phi))
\end{equation}
for any $z\geq0$, $c\geq 0$ and $\lambda\geq0$.
It follows from Theorem \ref{And1-9} and Theorem \ref{Reduction.Th2} that
\begin{equation}\label{15-1002}
R(1;c,\lambda)=c+\bar 
\rho(1;c,\lambda) 
\end{equation}
for all $c>0$ and $\lambda>0$. 

The following Lemma shows that  \eqref{15-1002}  is in fact a particular case of a much more general relationship.
\begin{lemma}\label{17-1105} For all $z>0$, $c> 0$ and $\lambda> 0$
\begin{equation}\label{17-1316}
R(z;c,\lambda)=c+\bar \rho(z;c,\lambda). 
\end{equation}
\end{lemma}
{\bf Proof of Lemma \ref{17-1105}}
It follows from \eqref{15-1002}  that
\begin{eqnarray}\label{31-1627}
R(z;c,\lambda)&=&z\inf_{(\psi,\phi)\in \mathscr{S}_0}
(\frac{c}{z}E_0\tau_\psi+\frac{\lambda}{z}\alpha(\psi,\phi)+\beta(\psi,\phi))\nonumber\\
\label{06-1000}
&=&z(c/z+\bar 
\rho(1;c/z,\lambda/z))=c+z\bar \rho(1;c/z,\lambda/z). 
\end{eqnarray}
Following the definitions of \eqref{05-2005} - \eqref{08-0959} and \eqref{08-1115}, it is not difficult to see  that
\begin{equation}\label{06-0954}
z\bar\rho(1;c/z,\lambda/z)=\bar\rho(z;c,\lambda)\;\mbox{for all}\;z,c,\lambda>0.
\end{equation}
Applying \eqref{06-0954} on the right-hand side of \eqref{06-1000}, we have \eqref{17-1316}
.$\Box$
%In fact, \eqref{17-1316} is still valid for $z=0$ and/or  for $\lambda=0$, being the respective value on both sides equal to $c$.

\begin{lemma}\label{17-1448}
%
%For any fixed $z\geq 0$ and $\lambda\geq 0$ it holds $R(z;0,\%lambda)=0$ and $\lim_{c\to 0}R(z;c,\lambda)=0$.
Function $R(z;c,\lambda)$ defined by  (\ref{17-1048}) is concave and jointly continuous on $\{z\geq 0,c\geq 0,\lambda\geq 0\}$.
\end{lemma}
{\bf Proof of Lemma \ref{17-1448}}
Because $R$ is defined as an infimum of a family of linear functions (see \eqref{17-1048}), it is concave on $\{z\geq 0,c\geq 0, \lambda\geq 0\}$, and it  follows from Theorem 10.1 in \cite{Rockaf}
that it is (jointly) continuous on $\{z>0,c>0,\lambda>0\}$.

It remains to show that
$R(z;c,\lambda)$ {is continuous at any point with }$ c=0,\lambda=0$ or $z=0$.

The most non-trivial part is when $c=0$ being $\lambda>0$ and $z>0$.

 Let $k$ be any natural number.  Let us define the stopping rule $\psi$ in such a way that 
$s_n^{\psi}=1$ for all $n\in \mathscr G^k$,  and let $\phi$ be a decision rule with
$\phi_n=I_{\{z_n\geq \lambda/ z\}}$, $n\in \mathscr G^k$. Then, by the Markov inequality, 
\begin{equation}\label{07-0959}
\lambda\alpha(\psi,\phi)=\lambda\sum_{n\in\mathscr G^k}p(n)P_{0}(Z_n\geq  \lambda/z)\leq   \sqrt{\lambda  z}\sum_{n\in\mathscr G^k}p(n)E_0Z_n^{1/2} =\sqrt{\lambda z}r^k,
\end{equation}
where
$$ r=\sum_{n\in\mathscr G}p(n)E_0Z_n^{1/2}<1
$$
(this latter inequality holds by virtue of  the well-known fact that the Hellinger divergence $E_0Z_1^{1/2}=\int f_{0}^{1/2}f_{1}^{1/2}d\mu<1$ whenever \eqref{09-1115} is satisfied).

Analogously, \begin{equation}\label{08-0830}z\beta(\psi,\phi)=z\sum_{n\in\mathscr G^k}p(n)P_{1}(Z_n^{-1}> (\lambda/z)^{-1})\leq 
\sqrt{\lambda z} \sum_{n\in\mathscr G^k}p(n)E_1Z_n^{-1/2}=
\sqrt{\lambda z}r^k.\end{equation}

Let now $\epsilon$ be any positive number. Taking  $c=0$ in \eqref{17-1048}
and 
\begin{equation}\label{08-1731}
k>\left(\ln \epsilon-\ln \big(\sqrt{\lambda z}\; \big)\right)/\ln r
\end{equation}
we see from \eqref{07-0959} and \eqref{08-0830} that $$R(z;0,\lambda)\leq \lambda\alpha(\psi,\phi)+z\beta(\psi,\phi)<2\epsilon.$$ Because $\epsilon$ is arbitrarily small, it follows
now that $R(z;0,\lambda)=0$, whatever $z>0,\lambda>0$.

To prove the continuity at any point $(z,0,\lambda)$, with $\lambda,z>0$, let us start with some $\lambda_n\to \lambda$, $z_n\to z$, $c_n\to 0$, as $n\to \infty$.

Let  $\epsilon>0$ be an arbitrary number again.
If  $k$ is large enough to satisfy \eqref{08-1731},
we have   $$R(z_n;c_n,\lambda_n)\leq c_nk+\lambda_n\alpha(\psi,\phi)+z_n\beta(\psi,\phi) <5\epsilon$$
for all $n$ such that $|\lambda_n-\lambda|<\epsilon$,
$|z_n-z|<\epsilon$, and $ c_n<\epsilon/k$,
that is,\\ $\lim_{n\to \infty} R(z_n;c_n,\lambda_n)=R(z;0,\lambda)=0$.

 Now, if  $\lambda=0$ and $z\geq 0$, $c\geq 0$, with the same definition of $(\psi,\phi)$ as above, and $k=1$,  we have that $\phi_{ n}=I_{\{z_n\geq 0\}}=1$, so $\alpha(\psi,\phi)=1$ and $\beta(\psi,\phi)=0$, thus, $R(z;c,0)=\inf_{(\psi, \phi)\in \mathscr{S}_0}(cE_0\tau_\psi+z\beta(\psi, \phi))\leq c$. On  the other hand $E_0\tau_\psi\geq 1$, so $R(z;c,0)=c$. If now  $(z_n, c_n, \lambda_n)\to (z, c,0)$, as $n \to \infty$, then $R(z_n;c_n,\lambda_n)\leq c_n+\lambda_n\to c$ as $n\to\infty$. On the other hand, $R(z_n;c_n,\lambda_n)=\inf_{(\psi, \phi)\in \mathscr{S}_0}(c_nE_0\tau_\psi+\lambda_n\alpha(\phi,\psi)+z_n\beta(\psi, \phi))\geq c_n\to c$, as $n\to\infty$.
 
 The case $z=0$ can be treated analogously.
$\Box$

In view of \eqref{17-1316}, it is convenient to extend the  definition of  $\bar\rho(z)=\bar\rho(z;c,\lambda)$ (defined initially for positive $c,\lambda$)  in such a way that \eqref{17-1316} hold for all  $z\geq 0,c\geq 0,\lambda\geq 0$, that is, defining $\bar\rho(z;c,\lambda)=0$ whenever any of its arguments is 0. Let us do so.

It follows now from Lemmas \ref{17-1105} and \ref{17-1448}
 
\begin{corollary}\label{29-1617}
The function $\bar \rho(z;c,\lambda)$
is concave and continuous on $\{z\geq 0,\,c\geq 0,\, \lambda\geq 0 \}$. 
\end{corollary}

The following Lemma is almost obvious, but will be very useful  in what follows.
\begin{lemma}\label{a0606}
 Let $F(z)$ be any  convex non-negative function on $\{z\geq 0\}$ such that $F(0)=0$. Then 
 \begin {equation}\label{a210624}F(z_1)\leq F(z_2)\end{equation} for all $0\leq z_1<z_2$, and the inequality in \eqref{a210624} is strict whenever  $F(z_1)>0$.
\end{lemma}
 {\bf Proof of Lemma \ref{22-161}}
	For any  $c>0$ and  $ \lambda>0$ let us define the functions  $D_1(z)=D_1(z;c, \lambda)=z-\bar{\rho}(z;c,\lambda)$ and $D_2(z)=D_2(z;c, \lambda)=\lambda-\bar{\rho}(z;c,\lambda)$, $z\geq 0$. Obviously, $D_1(\lambda)=D_2(\lambda)$.
	By virtue of the properties of $\bar{\rho}(z; c, \lambda)$, we  have that $D_1(z)$ and $D_2(z)$ are convex and continuous functions on $[0, \infty)$.

 By Jensen's inequality,
	\begin{equation}\label{Inequality-rho}
		\bar{\rho}(z)=\sum_{n\in \mathscr{G}}p(n) E_{0}\rho \big(zz_n\big) \leq  \sum_{n\in \mathscr{G}}p(n)\rho(z)=\rho(z)\leq g(z),
	\end{equation}
	so
	   $ g(z; \lambda)-\bar{\rho}(z;c,\lambda)=\min{\{ D_1(z), D_2(z) \}}\geq 0$.  

 By virtue of Lemma \ref{a0606}, if a positive $c$ is such that $c<D_1(\lambda)$ (which is equivalent to $\lambda>c+  \bar{\rho}( \lambda;c,\lambda)$, i.e. \eqref{22-161an}), then  there exists a unique $A<\lambda$ such that $D_1(A)=c$, and $D_1(z)<c$ for $z<A$ and 
  $D_1(z)>c$ for $z>A$.

 It is not difficult  to see that \eqref{18-1107} implies that $\lim_{z\to\infty}D_2(z)=0$. 
 
 Quite analogously to the above property of $D_1$, if $0<c<D_2(\lambda)=D_1(\lambda)$, then there exists a unique $B>\lambda$ such that $D_2(B)=c$, and $D_2(z)>c$ for $z<B$ and 
  $D_2(z)<c$ for $z>B$.
 
 Thus,   \eqref{22-1619}, \eqref{22-1632} and \eqref{22-1634} follow.

\begin{lemma}\label{27-0932}
  Let $A$ and $B$ be such that $0<A<B<\infty$. Then there exist $\lambda$ and $c$, such that $A\leq \lambda< B$ and $c>0$ and
  \eqref{22-1619} is satisfied.
\end{lemma}
{\bf Proof of Lemma \ref{27-0932}}
 Let $A$ and $B$ be such that $0<A<B<\infty$. For any
 $\lambda\in[A,B]$ let us define $c=c(\lambda)$ as a
solution to the equation
\begin{equation}\label{22}
c+\bar \rho(A;c,\lambda)=A.
\end{equation}
The existence of a unique solution $c=c(\lambda)$ of \eqref{22} follows  from the fact that the left-hand side of \eqref{22} is  a continuous strictly increasing function of $c$ taking values from 0 (at $c=0$) to $\infty$.
 Furthermore, $c(\lambda)$ is a continuous
function of $\lambda$, as an implicit function (\ref{22}) defined by
a function which is continuous in all its variables (by Corollary {\ref{29-1617}}).
In addition,  $c(\lambda)> 0$ for all $\lambda \in[A,B]$  because the contrary would imply, by virtue of  \eqref{22}, that $\bar\rho(A;0,\lambda)=A$, i.e. $A=0$, a contradiction.

Let us define now
$$
G(\lambda)=\lambda-\bar \rho(B;c(\lambda),\lambda)-c(\lambda),
$$
which is a continuous function of $\lambda$ as a composition of two
continuous functions.

Let us  show that
\begin{equation}\label{23}G(A)\leq 0,\quad\mbox{and}\quad  
G(B)>0.\end{equation}

Indeed, \begin{equation}\label{23an} G(A)=A-\bar \rho(B;c(A),A)-c(A)\leq  A-\bar \rho(A;c(A),A)-c(A)=0\end{equation}
(by
(\ref{22})).

Let us show now that
\begin{equation}\label{24}
G(B)=B-\bar \rho(B;c(B),B)-c(B)> 0.
\end{equation}
Taking into account that, by (\ref{22}),
$$
c(B)+\bar \rho(A;c(B),B)=A,
$$
we see that (\ref{24}) is equivalent to
\begin{equation}\label{a210622a}
B-\bar \rho(B;c(B),B)> A-\bar \rho(A;c(B),B).
\end{equation}
Because $F(z)=z-\bar \rho(z;c(B),B)$ satisfies the conditions of Lemma \ref{a0606}, the contrary to \eqref{a210622a} would imply that
$F(A)=0$, that is, $ A-\bar \rho(A;c(B),B)=0$, or, in view of \eqref{22}, $c(B)=0$, a contradiction. 
%But this is true due to the fact that $z-\bar \rho(z;c(B),B)$ is strictly
%increasing, at least in a neighbourhood  of $z=B$: indeed, it is an
%obviously convex non-negative function, taking value 0 at $z=0$, so there is a $z_0\geq 0$
%such that it is strictly increasing for $z\geq z_0$. Here $z_0$ can
%not be greater than or equal to $B$ because otherwise $\bar \rho(z;c(B),B)$
%would coincide with $g(z;B)$ for all $z\geq 0$, which is only
%possible when $f_{0}=f_{1}$ $\mu$-a.e., therefore contradicts \eqref{09-1115}.

Thus, (\ref{23}) is proved, so there exists $\lambda\in[A,B)$ such
that $G(\lambda)=0$, that is, we found $\lambda\in[A,B)$ and
$c=c(\lambda)>0$ such that
$$
c+\bar \rho(A;c,\lambda)=A\quad\mbox{and}\quad c+\bar \rho(B;c,\lambda)=\lambda
$$
which is equivalent to \eqref{22-1619}. $\Box$

{\bf Proof of Theorem \ref{t6}}
It follows from Lemma \ref{27-0932} by virtue of Lemma \ref{22-161}  that Condition $\mathfrak S_\infty(\psi)$ is satisfied.
Condition $\mathfrak D(\psi,\phi)$ is also satisfied because $B\geq \lambda$.
To apply Corollary \ref{And2-1} it remains to show that $(\psi,\phi)\in \mathscr S_0$, i.e. that
\begin{equation}\label{12-1534}
\sum_{n\in \mathscr G^k}p(n)E_0t_n^\psi=P_0(\tau_\psi\geq k)\to 0,\quad\mbox{as}\quad k\to\infty. 
\end{equation}
 It follows from Theorem 3.1 of \cite{Mukhopadhyay} that there exist $a>0$ and $0<r<1$
such that
$P_0(\tau_\psi\geq k)\leq ar^k$ for all natural $k$, so that  \eqref{12-1534} follows. The conditions of   
Theorem 3.1 of \cite{Mukhopadhyay} are satisfied because of our model assumtions 
and \eqref{09-1115}.

Hence, by Corollary  \ref{And2-1},
$(\psi,\phi)$ has a minimum value of  $K_0(\psi)$ among all tests 
$(\psi^\prime,\phi^\prime)$ satisfying the restrictions  \eqref{And2-3}
on the error probabilities.

To prove that the same test minimizes $K_1(\psi)$, we can apply Lemma \ref{27-0932} again,
just interchanging the hypothesized distributions.
Let us consider two simple hypotheses $H_0^\prime:$ ``the true distribution is 
given by $f_1$" vs. $H_1:$ ``the distribution corresponds to $f_0$".

In a very natural way, any test  $(\psi,\phi)$ from \eqref{13-0929} and \eqref{13-1011} 
is immediately 
adapted to the problem
of testing $H_0^\prime$ vs. $H_1^\prime$:% let $\psi_n^*=\psi_n^*(z_n^{-1)}=\psi_n(z_n)$ 
\begin{equation}\label{13-1633}
\psi_n^*(x^n)=\psi_n(x^n), \quad\mbox{and}\quad \phi_n^*(x^n)=1-\phi_n(x^n),
\end{equation}
so that
\begin{equation}
I_{\{  z_n^{-1}\in(B^{-1},A^{-1} ) \}}\leq 1-\psi_n^*\leq I_{\{  z_n^{-1}\in 
[B^{-1},A^{-1}] \}},
    \end{equation}
and 
\begin{equation}
\phi_n^* = I_{\{z_n^{-1}> B^{-1}\}}
\end{equation}
for all $n\in \mathscr G^k$ and $k=1,2,\dots$.

Let $\alpha^\prime$ and $\beta^\prime$  denote the  error probabilities in the problem of testing $H_0^\prime$
vs. $H_1^\prime$ and $K_0^\prime$ the average cost of observations, under $H_0^\prime$. Then, obviously,
\begin{equation}
 \alpha^\prime(\psi^*,\phi^*)=\beta(\psi,\phi), \quad\mbox{and} \quad \beta^\prime(\psi^*,\phi^*)=\alpha(\psi,\phi),
\end{equation}
and $K_0^\prime(\psi^*)=K_1(\psi)$.

Applying now Lemma \ref{27-0932} to the problem of testing $H_0^\prime$ vs. $H_1^\prime$ 
in the same way we applied it above for testing $H_0$ vs. $H_1$, we get, by Corollary  \ref{And2-1}, that $(\psi,\phi)$
minimizes $K_1(\psi)$ among the tests $(\psi^\prime,\phi^\prime)\in \mathscr S_1$ satisfying 
\eqref{And2-3}.

\newpage
\bibliography{document.bib}
\end{document}